\documentstyle[11pt,moriond,graphicx]{article}

\def\Journal#1#2#3#4{{#1} {\bf #2}, #3 (#4)}

\def\be{\begin{equation}}
\def\ee{\end{equation}}
\def\bea{\begin{eqnarray}}
\def\eea{\end{eqnarray}}

\begin{document}
\vspace*{4cm}
\title{RECENT CLEO $\Upsilon({\rm 1S})$ and $\Upsilon(\rm 3S)$ RESULTS}

\author{ T.E. COAN }

\address{Department of Physics, Southern Methodist University,\\
Dallas, TX, 75275, USA}

\maketitle\abstracts{Recent CLEO results 
on the observation of the $\Upsilon(1D)$ state
and a search for the $\eta_b(1S)$ state
from a data sample of $4.7 \times 10^6\,\Upsilon(3S)$ decays collected
with the CLEO-III detector, and a measurement of the $\eta^{\prime}$
production spectrum using $1.9 \times 10^6\,\Upsilon(1S)$ decays
collected with the CLEO-II detector are described.}

\section{Introduction}
CLEO is concluding a 1-year program to collect $\sim4\,{\rm
fb^{-1}}$ of integrated luminosity distributed over the $\Upsilon(1S),
\Upsilon(2S)$ and $\Upsilon(3S)$ resonances. These datasets will
augment the world supply of $b\bar{b}$ quark-antiquark bound state
data for these resonances by more than an order of magnitude. The data
is produced by the symmetric $e^+e^-$ collider CESR and collected by
the CLEO-III detector, a configuration of the CLEO detector with
excellent electromagnetic calorimetry and enhanced particle
identification capabilities.


\section{$\Upsilon(1D)$ Observation}

Heavy quarkonia (e.g., $b\bar{b}$ quark pairs) are an appealing
physical system for probing the strong interactions. Bound states
below open flavor threshold are narrow, leading to long lifetimes and
little mixing between excitation levels. The constituent quarks are
essentially non-relativistic, making the bound state system
particularly appropriate for analysis by lattice QCD calculations and
by effective theories via potential models.

We present evidence for the observation of the $\Upsilon(1D)$ state,
the first observation of a narrow $b\bar{b}$ bound state since the
early 1980's, using the following radiative 5-stage cascade decay that
ultimately produces 4 photons and a lepton pair in the final state:

\bea
\Upsilon(3S) & \rightarrow&  \gamma\,\chi_b(2P_J)\nonumber\\
\chi_b(2P_J) &\rightarrow& \gamma\,\Upsilon(1D)\nonumber\\
\Upsilon(1D)& \rightarrow& \gamma\,\chi_b(1P_J)\nonumber\\
\chi_b(1P_J) &\rightarrow& \gamma\,\Upsilon(1S)\nonumber\\
\Upsilon(1S) &\rightarrow& e^+e^-,\mu^+\mu^- 
\label{eq:chain}
\eea

The theoretical product branching fraction~\cite{rosner1} for this
sequential decay is $3.76\times 10^{-5}$. The dominant background
reaction is $\Upsilon(3S)\rightarrow \pi^0\pi^0\Upsilon(1S);
\Upsilon(1S)\rightarrow e^+e^-,\mu^+\mu^-$. An additional background
mode, the same cascade chain as Eq.~\ref{eq:chain} except that the
intermediate $\Upsilon(1D)$ is replaced by the $\Upsilon(2S)$ state,
must also be rejected.

\begin{figure}[h]
\hspace{4.5cm}
\includegraphics[scale=.33]{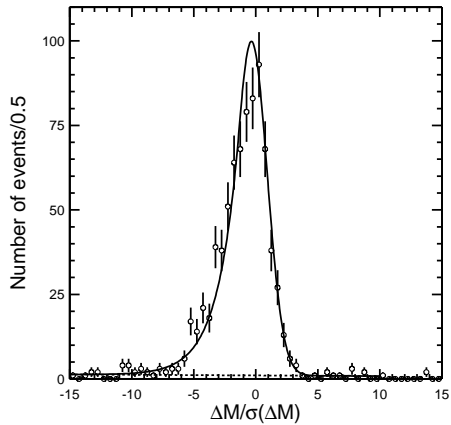}
\caption{Distribution of the recoil mass deviation from
the $\Upsilon(3S)$ mass for $\Upsilon(3S)\rightarrow \pi^0\pi^0\Upsilon(1S)$
candidate events. Dots are data, solid line is fit.\hfill
\label{fig:deltam}}
\end{figure}

Measuring the branching fraction for the background
$\Upsilon(3S)\rightarrow \pi^0\pi^0\Upsilon(1S)$ transitions is a
useful check of the technique used to reconstruct our cascade signal
mode.  A key analysis variable $\chi^2_{\pi^0\pi^0}$ used to
reconstruct $\Upsilon(3S)\rightarrow
\gamma\gamma\gamma\gamma\Upsilon(1S)$ decays is the minimal $\pi^0$
mass deviation chi-squared formed from among the three possible photon
pairs formed from four individual photons.  After cutting on
$\chi^2_{\pi^0\pi^0}$, the ratio $\Delta M/\sigma(\Delta M)$ is formed
from the deviation of the four-photon recoil mass from the
$\Upsilon(1S)$ mass $\Delta M = M_{recoil\,4\gamma}-M_{\Upsilon(1S)}$
and its expected mass resolution $\sigma(\Delta M)$. The distribution
$\Delta M/\sigma(\Delta M)$ is shown for data in Fig.~\ref{fig:deltam}
fit with a Monte Carlo determined shape. Combining the yield of
$737\pm28$ events with a 13.6\% Monte Carlo determined efficiency
implies a branching fraction
${\cal{B}}(\Upsilon(3S)\rightarrow\pi^0\pi^0\Upsilon(1S))
=(2.33\pm0.09\pm0.16)$\%, consistent with, but more precise than,
previous measurements~\cite{cleo2}${}^{,\,}$\cite{cusb}.

\begin{figure}[ht]
\hspace{1.25cm}
\includegraphics[scale=0.33]{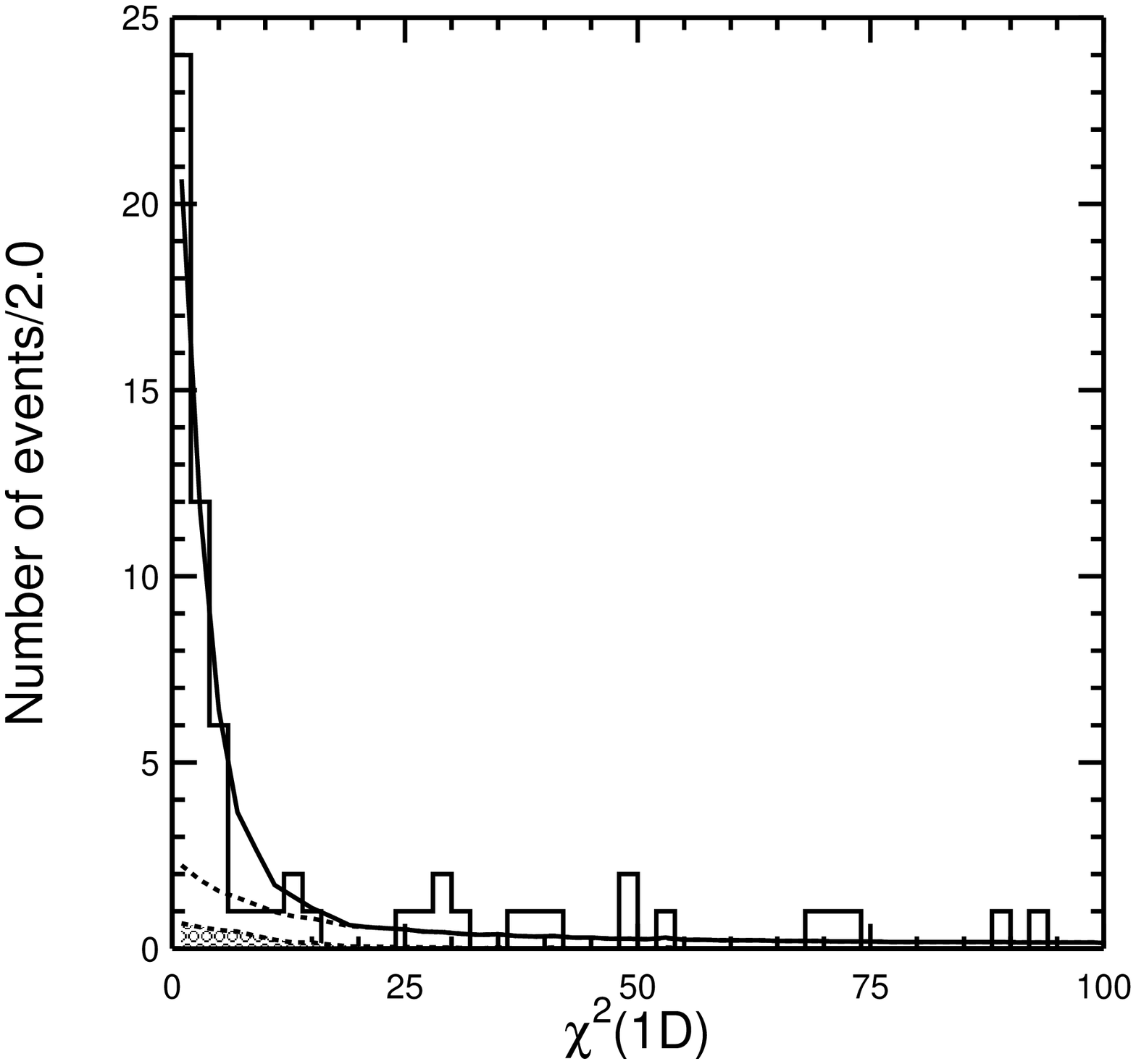}
\hspace{1.cm}
\includegraphics[scale=.33]{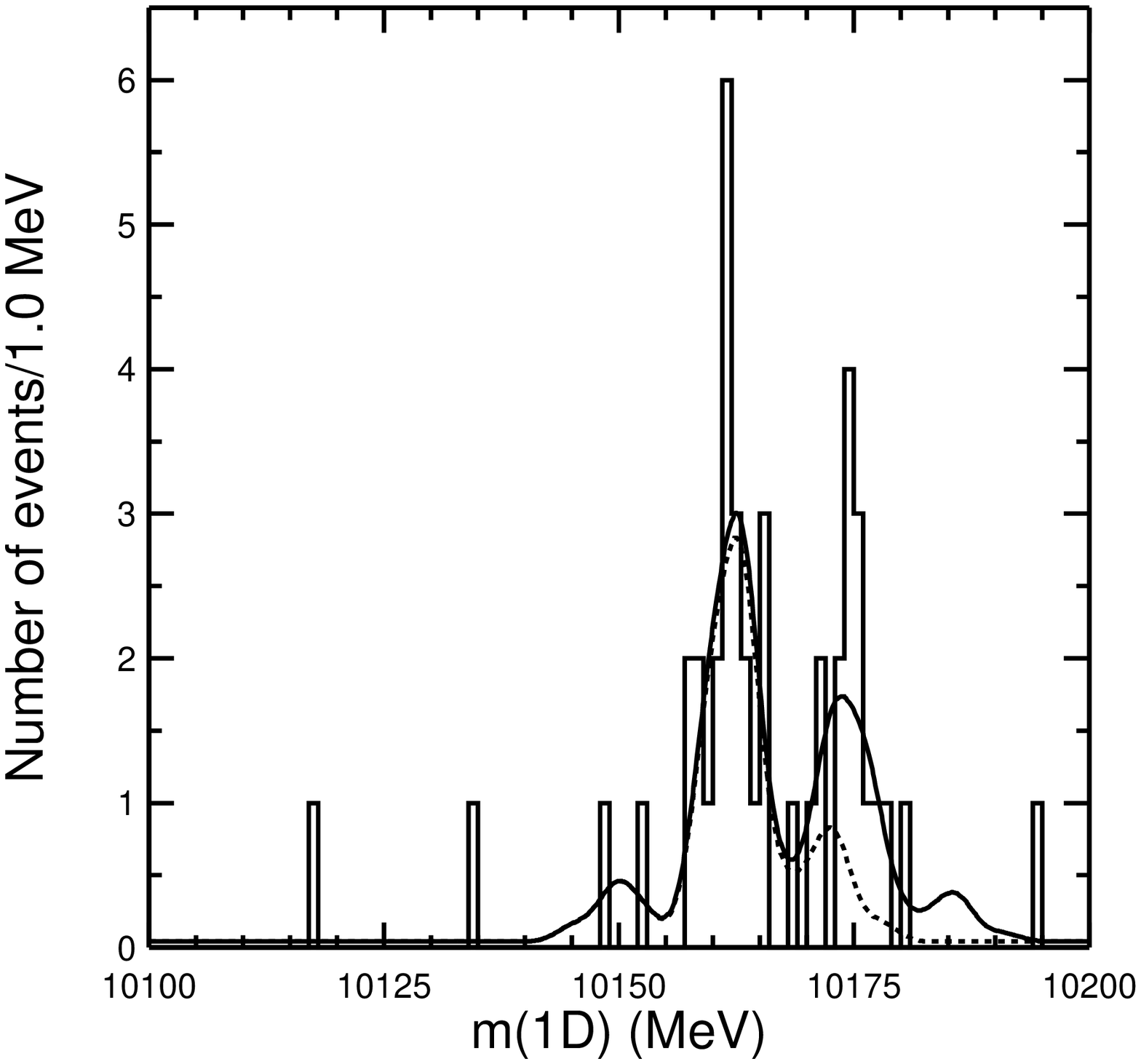}
\caption{The left hand side is a fit to the $\chi^2_{1D}$ for data. The
solid line is a fit to the signal contribution on top
of backgrounds. The right hand side is a fit of a two-peak structure
to the most-likely mass taken from the signal region of the left hand plot.\hfill
\label{fig:chi1d}}
\end{figure}

Confident that our technique for reconstructing multiple photons in a
single event is robust, we select $\Upsilon(1D)$ events by
constructing a chi-squared based variable $\chi^2(1D)$ that is a
function of the unknown $\Upsilon(1D)$ mass and the total angular
momentum $J$ of the $\chi_b$ states on either side of the
$\Upsilon(1D)$ in the cascade chain. Other chi-squared based variables
are used to reject background. Surviving events are shown in the left
hand side of Fig~\ref{fig:chi1d} as the histogram with the signal
region $\chi^2(1D)\leq 10$.  The fit of the signal on top of the
background is shown as the heavy and dashed lines, respectively,
producing a significance of 9.7$\sigma$.

The $\Upsilon(1D)$ mass is estimated by plotting for each event in the
signal region of the left hand plot in Fig.~\ref{fig:chi1d} the value
of the mass that minimizes $\chi^2(1D)$. Monte Carlo studies predict a
peak at the true mass plus two satellite peaks due to
mis-reconstructed photons from shower leakage in the calorimeter. The
mass distribution for data is shown as the histogram in the right hand
plot of Fig.~\ref{fig:chi1d}. Although the $\Upsilon(1D)$ can be
produced in any of the $J=1,2,3$ states, our selection efficiency is
largest for the $\Upsilon(1D_1)$ and $\Upsilon(1D_2)$ states while
theory predicts that the $\Upsilon(1D_2)$ state should be
preferentially produced in the decay cascade. Fitting the histogram
with two peaks, each with two satellites, produces a peak at
$M_{low}=10161.2\pm0.7\,{\rm MeV/c^2}$ and at
$M_{high}=10174.2\pm1.3\,{\rm MeV/c^2}$, with an overall 58\%
C.L. Fitting under the assumption that there is no low mass peak
produces a 0.04\% C.L. and the difference in significance between
these two fits implies a significance of the low mass peak of
$6.8\sigma$. Fitting the histogram with just a single peak plus two
satellites yields a 43\% C.L. fit with a peak at $10162.0\pm0.5\,{\rm 
MeV/c^2}$.

Averaging over the different fits yields a $\Upsilon(1D)$ mass of
$10162.2\pm1.6\,{\rm MeV/c^2}$ with the $J=2$ state favored, but the $J=1$ state
not excluded. Based on our measured $\Upsilon(1D)$ yields and Monte
Carlo determined efficiency, the preliminary overall branching
fraction for the 5-stage decay cascade of the $\Upsilon(3S)$ via the
$\Upsilon(1D)$ is $(3.3\pm0.6\pm0.5)\times10^{-5}$.

\section{Search for $\eta_b(1S)$}

The spin-singlet state of the $b\bar{b}$ system, including the
$b\bar{b}$ ground state $\eta_b(1S)$, has yet to be observed.  The
reaction $e^+e^-\rightarrow\gamma^{\ast}\rightarrow b\bar{b}$ produces
$b\bar{b}$ states with the same $J^{PC}$ quantum numbers as the photon
so the $\eta_b(1{}^1S_0)$ state cannot be produced directly at CESR.
However, the state is accessible through decays from the spin-triplet
$\Upsilon(nS)$ states via magnetic dipole (M1) radiative transitions
which occur between states with opposite quark spin configuration but
the same orbital angular momentum. The rate for M1 transitions between
the $\Upsilon$ and the $\eta_b$ is given by~\cite{m1}

\be
\Gamma[\Upsilon(nS)\rightarrow \gamma\,\eta_b(n^{\prime}S)]\propto
I^2E_{\gamma}^3,
\label{eq:dipole}
\ee

\noindent where $n$ and $n^{\prime}$ are the principal quantum numbers
of the respective $b\bar{b}$ systems, $I$ is an overlap integral
between initial and final $b\bar{b}$ states and $E_{\gamma}$ is the
energy of the emitted photon.

With a data sample of $4.7\times10^6\,\Upsilon(3S)$ events, CLEO
searches for transitions with $n\neq n^{\prime}$ to avoid photon
energies in the $100\,$MeV range which are susceptible to background
contamination from $\pi^0$ decays. The inclusive photon search
technique for $\Upsilon(3S)\rightarrow \gamma\eta_b(1S)$ is tuned
using the electric dipole (E1) transitions $\chi_b(2P)\rightarrow
\Upsilon(1S)$ where the photon energy is $\sim 800\,$MeV, similar to
what is expected~\cite{m1} for the signal M1 transition.

\begin{figure}[htb]
\hspace{1.25cm}
\includegraphics[scale=.33]{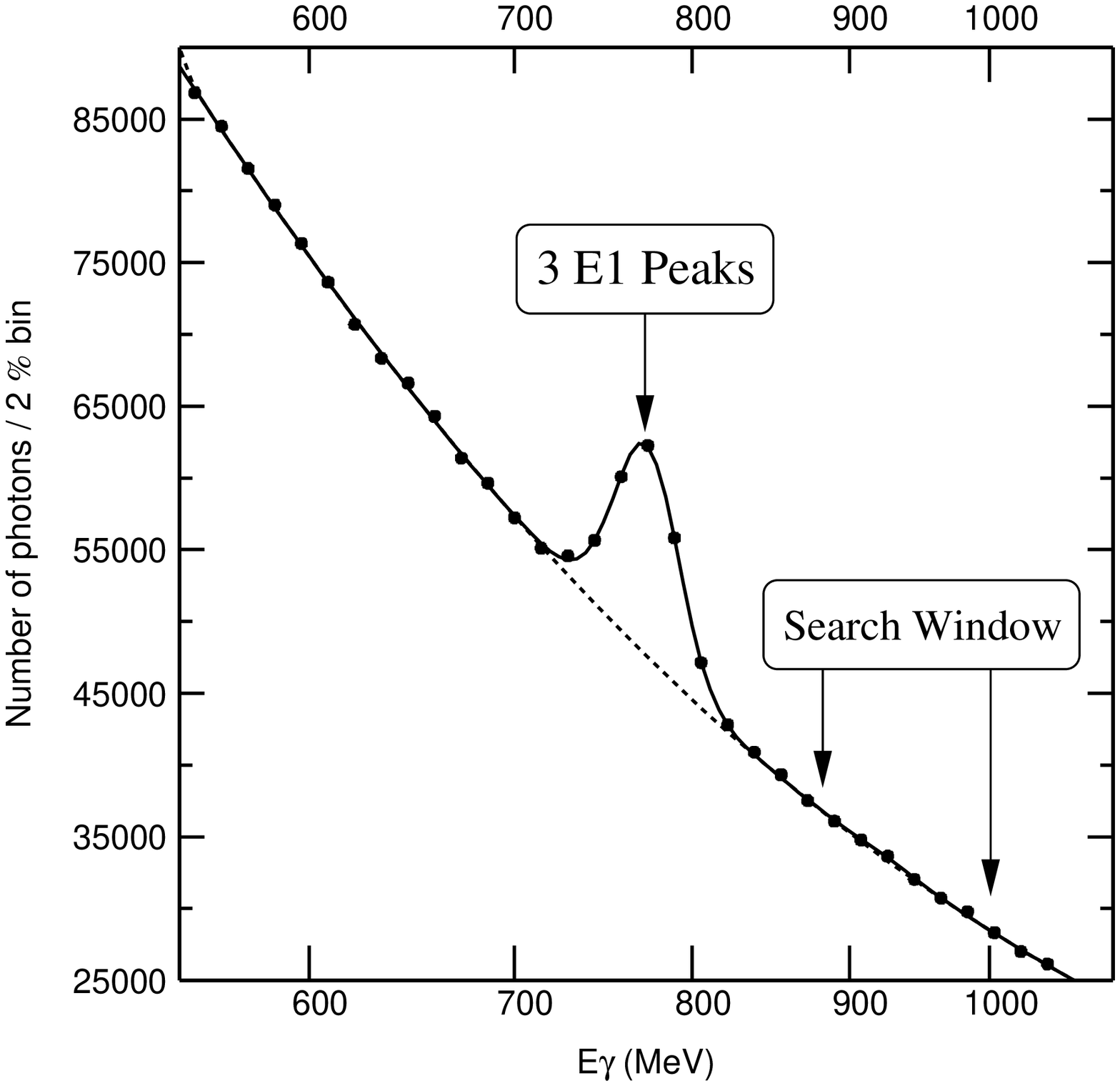}
\hspace{1.0cm}
\includegraphics[scale=0.33]{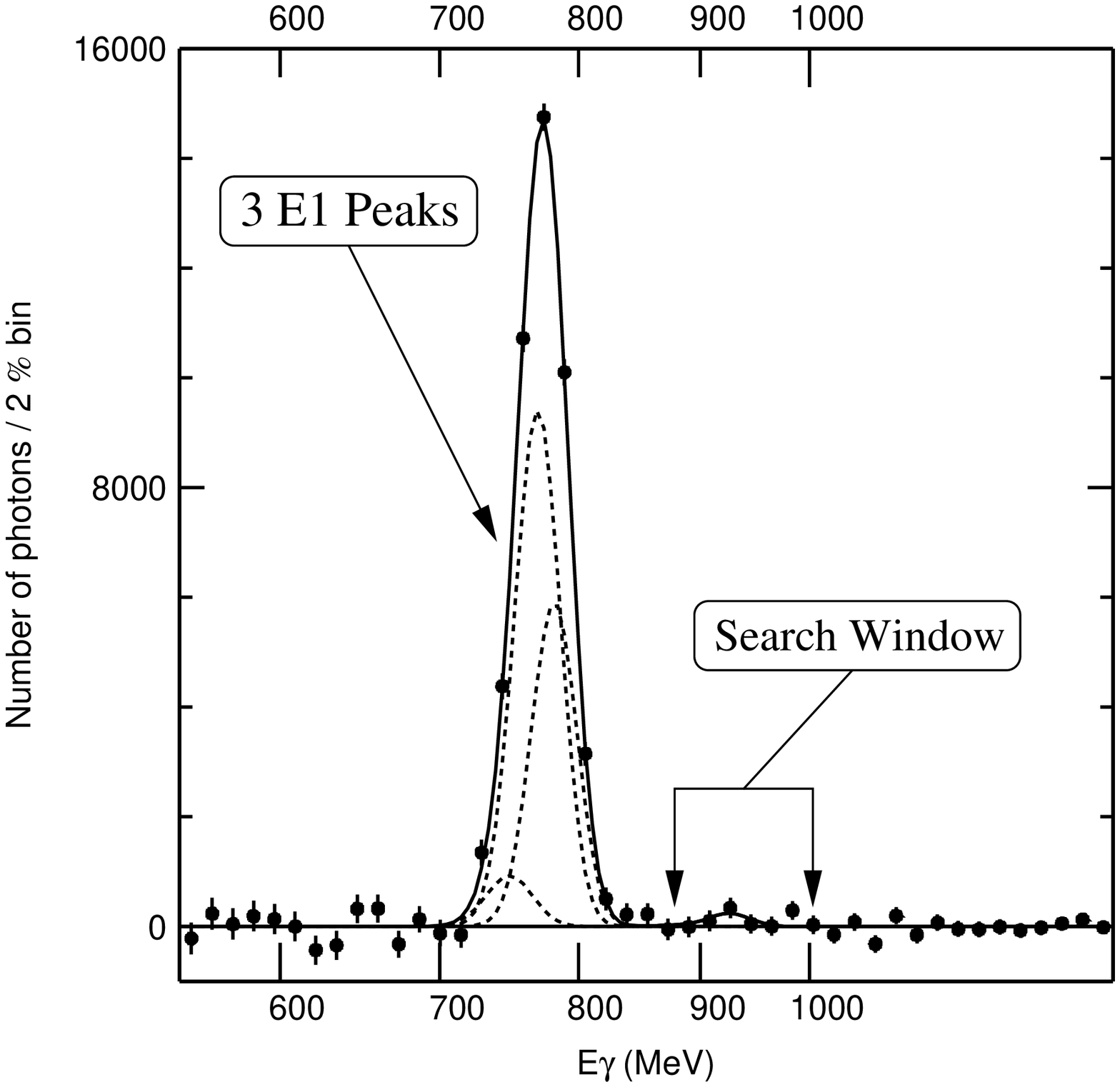}
\caption{The inclusive photon spectrum in the E1 peak and the $\eta_b$ search
region before (left) and after (right) background subtraction.\hfill
\label{fig:gammas}}
\end{figure}

The inclusive photon spectrum between approximately $550\,{\rm MeV}$
and $1\,{\rm GeV}$ is shown in the left hand side of
Fig.~\ref{fig:gammas} where the dots indicate data and the dashed line
is background.  A logarithmic energy scale is used to preserve energy
resolution. The central peak is a superposition of three gaussians for
the three $\chi_b(2P_J)$ E1 transitions and the search window is also
indicated.

\begin{figure}[ht]
\hspace{4.0cm}
\includegraphics[scale=0.42]{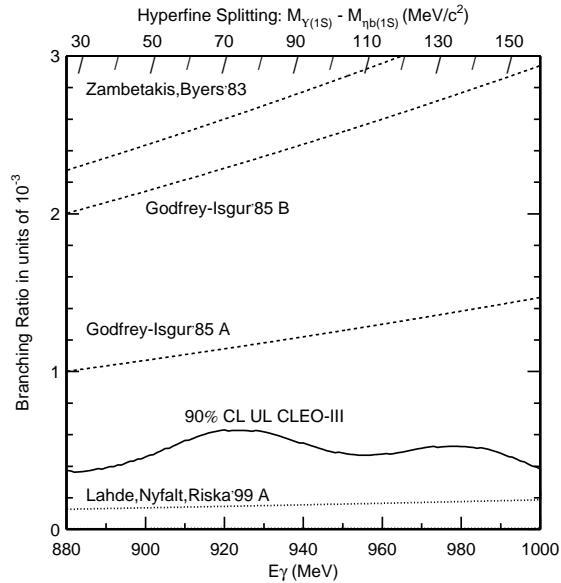}
\caption{Upper limits on ${\cal{B}}(\Upsilon(3S)\rightarrow \gamma\,\eta_b(1S))$
as a function of $E_{\gamma}$. Model predictions are from Ref.~[4].
\label{fig:exclude}}
\end{figure}

The right hand side of Fig.~\ref{fig:gammas} shows the background
subtracted central peak and the search region fitted with a Crystal
Ball lineshape. Multiple fits are performed for various photon
energies in the search region and no signal is seen. This lack of
signal is converted into a 90\% C.L. upper limit shown in
Fig.~\ref{fig:exclude}, excluding a variety of phenomenological
models~\cite{m1}.

\section{$\Upsilon(1S)\rightarrow \eta^{\prime}X$ Production}

\begin{figure}[h]
\hspace{5.5cm}
\includegraphics[scale=0.37]{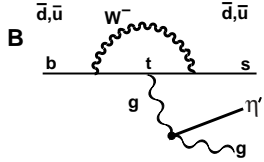}
\caption{Feynman diagram for $b\rightarrow sg\eta^{\prime}$.
\label{fig:gnn}}
\end{figure}

An unexpectedly large $B\rightarrow\eta^{\prime} X_s$ inclusive decay
rate in the momentum range $2\leq p_{\eta^{\prime}} \leq 2.7\,$ GeV/c
observed by CLEO~\cite{cleoetap} and BABAR~\cite{babaretap} is
possibly described by an anomalously large coupling between the
$\eta^{\prime}$ and two gluons~{\cite{2gluons}${}^{,}$
\cite{models}${}^{,\,}$\cite{2gluons2}}
in the underlying $b\rightarrow sg$ mechanism thought responsible for
$\eta^{\prime}$ production. Here the process $b\rightarrow sg$ is
followed by the two gluon coupling to the $\eta^{\prime}$ shown in
Fig.~\ref{fig:gnn}.

The effective $\eta^{\prime}g^{\ast}g$ coupling can be written as~\cite{workers}
\be
H(q^2)\epsilon_{\alpha\beta\mu\nu}q^{\alpha}k^{\beta}\epsilon_1^{\mu}
\epsilon_2^{\nu},
\ee

\noindent where $H(q^2)$ is the $\eta^{\prime}g^{\ast}g$ transition
form factor, $q=p_b-p_s$ is the four-momentum of the virtual hard
gluon ($g^{\ast}$) and $k$ is the momentum of the soft gluon
($g$). Workers~\cite{workers} have shown that the $q^2$ region
relevant to the process $b\rightarrow sg\eta^{\prime}$ is accessible
in high energy $\eta^{\prime}$ production in $\Upsilon(1S)$ decays so
that constraints can be placed on $H(q^2)$ from the fast
$\eta^{\prime}$ spectrum in $\Upsilon(1S)\rightarrow ggg$
decays. Various choices~\cite{models}${}^{,}$\cite{workers} are
available for $H(q^2)$.

We use $1.86\times 10^6\,\Upsilon(1S)$ decays collected with the old
CLEO~II detector and detect $\eta^{\prime}$ mesons through the decay
channel $\eta^{\prime}\rightarrow \eta\pi^+\pi^- (\eta\rightarrow
\gamma\gamma)$. 
This data sample can be divided into three parts:
\be
N_{all} = N_{\Upsilon(1S)\rightarrow ggg} +
N_{\Upsilon(1S)\rightarrow q\bar{q}} + 
N_{e^+e^-\rightarrow q\bar{q}\,},
\nonumber
\ee

\noindent where the last piece is the most problematic background to
correct for.  We do this by using $1.19\,{\rm fb^{-1}}$ of
off-resonance data collected just below the $\Upsilon(4S)$ resonance
by a process that maps the $\eta^{\prime}$ production spectra at the
higher $\Upsilon(4S)$ energy to the lower $\Upsilon(1S)$ energy by a
procedure analogous to what is done when one maps one normalized
probability distribution into another one with a different domain.
The remapping is done with the aid of the variable
$Z=E_{\eta^{\prime}}/E_{beam}$, the fractional energy of the
$\eta^{\prime}$.


\begin{figure}[h]
\hspace{4.5cm}
\includegraphics[scale=0.33]{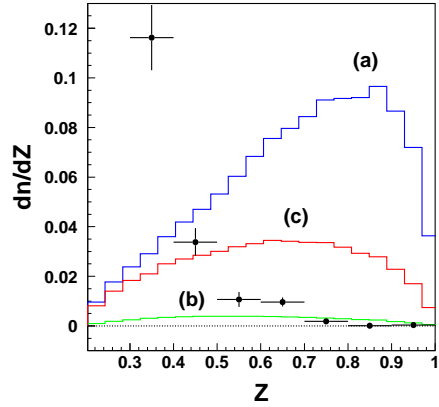}
\caption{Measured $dn/dZ$ spectrum of $\Upsilon(1S)\rightarrow ggg
\rightarrow\eta^{\prime}X$ compared with theoretical predictions. See text
for explanation.\hfill
\label{fig:kagan}}
\end{figure}

Defining three differential branching fractions $dn/dZ$
\bea
{dn(ggg)\over dZ} &=& {d{\cal{B}}(\Upsilon(1S)\rightarrow ggg\rightarrow
\eta^{\prime}X)\over dZ\times {\cal{B}}(\Upsilon(1S)\rightarrow ggg)}\nonumber\\
{dn(q\bar{q})\over dZ} &=& {d{\cal{B}}(\Upsilon(1S)\rightarrow q\bar{q}\rightarrow
\eta^{\prime}X)\over dZ\times {\cal{B}}(\Upsilon(1S)\rightarrow q\bar{q})}\nonumber\\
{dn(1S)\over dZ} &=& {d{\cal{B}}(\Upsilon(1S)\rightarrow \eta^{\prime}X)\over dZ}\nonumber
\label{eg:diff}
\eea

\noindent permits us to use the $Z$ spectrum of $\eta^{\prime}$ and
discriminate among the various phenomenological models that predict
different forms of $H(q^2)$. Fig.~\ref{fig:kagan} shows the comparison
between the measured $dn(ggg)/dZ$ spectrum (dots) and various
phenomenological predictions (histograms). Only the region $Z\ge0.7$
is relevant for comparison since the models require the
$\eta^{\prime}$ be ``fast.'' The model labeled ``b'' is representative
of perturbative QCD~\cite{ali}${}^{,\,}$\cite{muta} and fits the data
best. There is no evidence for any anomalously high
$\eta^{\prime}g^{\ast}g $ coupling at large $\eta^{\prime}$ energies.

\section*{Acknowledgments}
The author acknowledges support by the U.S. Department of Energy under
grant DE-FG03-95ER40908.

\end{document}